\documentclass[12pt]{article}
\usepackage{a4wide}
\usepackage{amssymb}
\usepackage{color}
\begin{document}
{\renewcommand{\thefootnote}{\fnsymbol{footnote}}
\begin{center}
{\LARGE  States in non-associative quantum mechanics:\\ Uncertainty relations
  and semiclassical evolution}\\
\vspace{1.5em}
Martin Bojowald,$^1$\footnote{e-mail address: {\tt bojowald@gravity.psu.edu}}
 Suddhasattwa Brahma,$^1$\footnote{e-mail address: {\tt sxb1012@psu.edu}} Umut B\"{u}y\"{u}k\c{c}am,$^1$\footnote{e-mail address: {\tt uxb101@psu.edu}} and Thomas Strobl$^2$\footnote{e-mail address: {\tt strobl@math.univ-lyon1.fr}}\\
\vspace{0.5em}
$^1$ Institute for Gravitation and the Cosmos,\\
The Pennsylvania State
University,\\
104 Davey Lab, University Park, PA 16802, USA\\
\vspace{0.5em}
$^2$ Institut Camille Jordan, Universit\'e Claude Bernard Lyon 1,\\
43 boulevard du 11 novembre 1918, 69622 Villeurbanne cedex, France
\vspace{1.5em}
\end{center}
}

\setcounter{footnote}{0}

\newtheorem{theo}{Theorem}
\newtheorem{lemma}{Lemma}
\newtheorem{defi}{Definition}

\newcommand{\proofend}{\raisebox{1.3mm}{\fbox{\begin{minipage}[b][0cm][b]{0cm}
\end{minipage}}}}
\newenvironment{proof}{\noindent{\it Proof:} }{\mbox{}\hfill \proofend\\\mbox{}}
\newenvironment{ex}{\noindent{\it Example:} }{\medskip}
\newenvironment{rem}{\noindent{\it Remark:} }{\medskip}

\newcommand{\case}[2]{{\textstyle \frac{#1}{#2}}}
\newcommand{\lP}{\ell_{\mathrm P}}

\newcommand{\md}{{\mathrm{d}}}
\newcommand{\tr}{\mathop{\mathrm{tr}}}
\newcommand{\sgn}{\mathop{\mathrm{sgn}}}

\newcommand*{\R}{{\mathbb R}}
\newcommand*{\N}{{\mathbb N}}
\newcommand*{\Z}{{\mathbb Z}}
\newcommand*{\Q}{{\mathbb Q}}
\newcommand*{\C}{{\mathbb C}}

\begin{abstract}
 A non-associative algebra of observables cannot be represented as operators
 on a Hilbert space, but it may appear in certain physical situations. This
 article employs algebraic methods in order to derive uncertainty relations
 and semiclassical equations, based on general properties of quantum moments.
\end{abstract}

\section{Introduction}

Quantum mechanics represents the classical Poisson algebra of basic
variables $q_j$ and $p_k$, $\{q_j,p_k\}=\delta_{jk}$, as an operator
algebra acting on a Hilbert space, so that the Poisson bracket is
turned into the commutator $[\hat{q}_j,\hat{p}_k]=i\hbar \delta_{jk}$
of basic operators.\footnote{This latter equation, as usual, holds on
a dense subspace only. From a mathematical perspective, it is
convenient to consider bounded operators obtained after exponentiating
$\hat{q}_i$ and $\hat{p}_j$, resulting in the Weyl algebra.  In the
present paper, however, we focus on conceptual questions of the
construction of non-associative quantum mechanics and related
consequences in possible physical applications, postponing more
mathematical issues to later work.}  The Jacobi identity satisfied by
a Poisson bracket has an analog in the associativity of the operator
product: A simple calculation shows that
\begin{equation} \label{JacAss}
 \epsilon^{ijk}[[\hat{O}_i,\hat{O}_j],\hat{O}_k]=2
 \epsilon^{ijk}[\hat{O}_i,\hat{O}_j,\hat{O}_k]
\end{equation}
where the 3-bracket on the right-hand side is used to denote the associator of
the product of the quantum observables,
$[\hat{O}_1,\hat{O}_2,\hat{O}_3]:=(\hat{O}_1\hat{O}_2)\hat{O}_3-
\hat{O}_1(\hat{O}_2\hat{O}_3)$. If the Jacobiator of the classical bracket ---
or the Jacobiator of the operator product introduced on the left-hand side of
(\ref{JacAss}) --- vanishes for all triples of operators $\hat{O}_i$, an
associative operator algebra is consistent with Dirac's basic quantization
rule relating Poisson brackets to commutators. (These concepts have been
formalized mathematically in different ways, for instance in the frameworks of
group-theoretical quantization \cite{Isham}, geometric quantization
\cite{Woodhouse}, and deformation quantization.)

For classical systems with modified brackets, such as twisted Poisson
structures \cite{TopoTwist,WZWTwist,Twisted}, the Jacobi identity may no
longer hold true and be replaced by a non-zero Jacobiator
$\epsilon^{ijk}\{\{O_i,O_j\},O_k\}\not=0$. As the quantum analog, there must
be a non-zero associator $[\hat{O}_1,\hat{O}_2,\hat{O}_3]\not=0$ of a
non-associative operator algebra. Such an algebra cannot be represented on a
Hilbert space in the standard way, and alternatives making use, for instance,
of non-associative $*$-products must be developed.  In this paper, we focus on
the general aspects of states on a non-associative operator algebra and see
how the basic notions familiar from quantum mechanics can be derived in
representation-independent terms. In some respects (and unless extra
conditions on states are imposed) the results seem to differ from existing
constructions using non-associative $*$-products
\cite{NonGeoNonAss,MSS1,BakasLuest,MSS2,MSS3}.

Non-associative structures have recently gained interest in the context of
certain flux compactifications of string theory and double-field theory
\cite{DualDoubled,NonAssGrav,NonGeoNonAss,TwistedNonAss,NonAssDef}. They have
played a role in the understanding of gauge anomalies, and also appear in
``standard'' quantum mechanics if a charged particle is coupled to a density
of magnetic monopoles \cite{MagneticCharge}. These monopoles need not be
fundamental, and therefore the systems may describe realistic physics in some
analog models of condensed-matter systems (see for instance
\cite{MonopoleIce}). A related version is realized in chiral gauge theories
\cite{CurrentAlgebra,CurrentCommutator,NonAssChiral}. We briefly review how
non-Poisson brackets or non-associative algebras appear, which will present
the main example to keep in mind throughout this article.

In the presence of a magnetic field with vector potential $A_i$, the canonical
momentum of a particle with mass $m$ and charge $e$ is $\pi_i=mv_i+eA_i$ in
terms of the velocity $v_i=\dot{q}_i$. While the momentum
components obey canonical Poisson brackets with the position variables and
have zero brackets with one another, the velocity components or the
kinematical momentum components $p_i=mv_i$ have brackets related to the
magnetic field:
\begin{equation} \label{vv}
 \{p_i,p_j\}= e \epsilon_{ijk}B^k\,.
\end{equation}
(We use the convention that repeated indices are summed over.)  These brackets
define a Poisson structure provided the magnetic field is divergence free.

If the divergence is non-zero, the magnetic field no longer has a vector
potential, but one may still use (\ref{vv}) as the definition of a bracket on
phase space (together with $\{q_i,p_j\}= \delta_{ij}$ and antisymmetry). One
then computes a non-zero Jacobiator
\begin{equation} \label{Jacob}
 \epsilon^{ijk} \{\{p_i,p_j\},p_k\}=-2e \partial_lB^l\,.
\end{equation}
The corresponding quantum mechanics cannot be represented by an associative
operator algebra acting on a Hilbert space.

For a constant monopole density, the bracket is twisted Poisson
\cite{TopoTwist,WZWTwist,Twisted} and can be realized as a Malcev algebra
\cite{Malcev,Malcev2}.  The $*$-product constructions of
\cite{NonGeoNonAss,MSS1,BakasLuest,MSS2,MSS3} simplify for a constant density
and allow several explicit results to be derived, but they hold more
generally. Our calculations here are complementary and allow for $\partial_l
B^l$ to be non-constant even in some explicit results. The existence of
relevant algebras and states based on our relations alone is more difficult to
show, but if they are assumed to exist, several properties can be derived
efficiently by considering expectation-value functions $\omega\colon{\cal
  A}\to{\mathbb C}$.

From the physical perspective, this example is of interest because the
existence of a magnetic monopole density, fundamental or effective,
{\it somewhere} within the system necessitates a modification of very
fundamental aspects of quantum mechanics. The notion of a Hilbert
space is a non-local one, for instance in the sense that
wave functions in the Schr\"odinger representation are normalized by
an integration over all of space. Nevertheless, for meaningful
experiments it must be possible to construct a local description of
quantum physics outside the magnetic monopole density, where it has to
reproduce the established and experimentally verified quantum
properties (at least to a very high precision). This (perhaps
hypothetical) physical system thus provides an interesting playground
and a test for the development of non-associative quantum mechanics.

\section{Properties of states}

As briefly derived in the example of a magnetic monopole density, we assume
that we have an algebra ${\cal A}$ of observables, which includes elements
$\hat{q}_i$ and $\hat{p}_j$ (as well as a unit ${\mathbb I}$) and obeys the
relations
\begin{eqnarray}
 [\hat{q}_i,\hat{q}_j]&=& 0 \label{qq}\\
 \left [\hat{q}_i,\hat{p}_j\right ]&=& i \hbar \delta_{ij} \\
 \left [\hat{p}_i,\hat{p}_j\right]&=&i \hbar  e \epsilon_{ijk}
 \hat{B}^k \label{pp}  \\
\left [ \hat{p}_i,\hat{p}_j, \hat{p}_k \right ] &=& -\hbar^2 e \epsilon_{ijk}
 \widehat{\partial_lB^l} \label{Asso}
\end{eqnarray}
where $i,j,k \in\{1,2,3\}$. Summation over double indices is assumed, and, as
before, the 3-bracket denotes the associator.  We further assume that, other
than (\ref{Asso}), all associators between the fundamental variables
$\hat{q}_i$ and $\hat{p}_j$ vanish.  Among basic operators, only the
associator (\ref{Asso}) is non-zero because of the non-vanishing Jacobiator
(\ref{Jacob}).

In (\ref{pp}) and (\ref{Asso}), $\hat{B}^k\in {\cal A}$ (and similarly
$\widehat{\partial_lB^l}$) is obtained by inserting $\hat{q}_i$ in the
classical function $B^k(q_i)$. Since the $\hat{q}_i$ commute and associate
with one another, $\hat{B}^k$ is well-defined for polynomial $B^k$. For
non-polynomial magnetic fields, we assume that $\hat{B}^k$ can be defined by a
formal power series.  With the magnetic field and also its divergence allowed
to be functions of $q_i$, the relations may be non-linear.  If the
magnetic-field components are assumed to be analytic functions, then also
their derivatives are well-defined, which we will use in some semiclassical
expansions. The assumption of analyticity may have to be weakened in some
physical situations because it is not consistent with a monopole density of
compact support. For the algebra, we need the first derivatives of $B^i$, so
that these functions should at least be differentiable.

At this point, we encounter the first existence question. In
associative cases, it is known that $\hat{q}_i$ and $\hat{p}_j$ are
not bounded in Hilbert-space representations. It is then more
convenient to use exponentiated (Weyl) operators for explicit
constructions of algebra representations. In the present case,
Hilbert-space representations cannot exist at all, and existence
questions are more complicated. In this article, we take a pragmatic
view and assume that an algebra with the relations
(\ref{qq})--(\ref{Asso}) (as well as a $*$-relation introduced below)
exists.  Our aim is to derive properties of states which are of
interest for physical questions and can be obtained using only the
given relations. This view is akin to the one taken in particle
physics, where it is difficult to show that interacting quantum field
theories do indeed exist, but powerful computational methods are still
available and can be compared with observations.

The relations (\ref{qq})--(\ref{pp}) are direct translations of basic
brackets, the first two of standard form and (\ref{pp}) derived from
(\ref{vv}). The non-zero Jacobiator (\ref{Jacob}) implies that there
must be a non-zero associator. However, (\ref{JacAss}) shows that only
the totally antisymmetric part of the associator is determined by the
correspondence between classical brackets and
commutators. Contributions to the associator which are not totally
antisymmetric can be considered as quantization choices, which one may
be able to choose so as to realize certain simplifications. For now,
we will assume simplifications which appear to be consistent with the
equations (\ref{qq})--(\ref{Asso}), postponing a more precise
construction of ${\cal A}$ to later work. 

We could assume the associator between any three elements of the algebra
${\cal A}$ to be completely antisymmetric, or equivalently
\begin{eqnarray}\label{alternating}
A(BB) &=& (AB)B\\
(AA)B &=& A(AB)\\
(AB)A &=& A(BA)
\end{eqnarray}
for all $A,B,C\in{\cal A}$. Any algebra satisfying these conditions (two of
which imply the third one) is called an \emph{alternative algebra}.  For such
an algebra, we have additional relations between algebra elements which are
not as strong as assocativity but will turn out to be useful: An alternative
algebra satisfies the Moufang identities \cite{Moufang}
\begin{eqnarray}
 C(A(CB)) &=& (CAC)B\\
 ((AC)B)C &=& A(CBC)\\
 (CA)(BC) &=& C(AB)C\,. \label{Moufang}
\end{eqnarray}
(If (\ref{alternating}) holds, we do not need to set further paranthesis in
(\ref{Moufang}).) These identities are also useful for an extension of some of
the measurement axioms of quantum mechanics to non-associative versions
\cite{OctQM}. The algebras constructed by $*$-products in
\cite{MSS1,BakasLuest,MSS2,MSS3}, with the same basic associator (\ref{Asso}),
are {\em not} alternative.\footnote{We are grateful to Peter Schupp and
  Richard Szabo for pointing this out to us.} Our explicit results derived in
the rest of this paper only require (\ref{Asso}) to be totally antisymmetric,
and the corresponding Moufang identity for $A$, $B$ and $C$ linear in the
$\hat{p}_i$. They will therefore also hold for the known $*$-algebra
realizations of (\ref{Asso}), but there may be deviations at higher moments or
$\hbar$-orders.

We turn ${\cal A}$ into a $*$-algebra by requiring $\hat{q}_i$ and
$\hat{p}_j$ to be self-adjoint. (We then have the usual relations,
such as $(\lambda \hat{p}_1)^* = \lambda^* \hat{p}_1$ for all
$\lambda\in {\mathbb C}$, and $(AB)^* = B^* A^*$ for all $A,B \in
{\cal A}$.) This requirement is consistent with (\ref{Asso}) thanks to
the alternative nature of the algebra: for self-adjoint
$\hat{p}_i^*=\hat{p}_i$, we then have
\begin{equation}
 [\hat{p}_1,\hat{p}_2,\hat{p}_3]^*=\hat{p}_3(\hat{p}_2\hat{p}_1)-
 (\hat{p}_3\hat{p}_2)\hat{p}_1= -[\hat{p}_3,\hat{p}_2,\hat{p}_1]=
 [\hat{p}_1,\hat{p}_2,\hat{p}_3]
\end{equation}
so that both sides of (\ref{Asso}) are self-adjoint.

\subsection{The Cauchy--Schwarz inequality and uncertainty relations}

In treatments of algebra theory relevant for quantum mechanics it is often
assumed that one is dealing only with associative algebras. Several important
results no longer apply in the non-associative case. However, a notable
exception is the Cauchy--Schwarz inequality. It is important in quantum
mechanics because it leads to the uncertainty relation, and fortunately, this
result is still available for non-associative algebras. Even the standard
proof can be used without modifications, which we sketch here for
completeness.

For any complex-valued, positive linear functional $\omega$ on the
algebra ${\cal A}$ (that is, $\omega(A^*A)\geq0$ for all $A\in {\cal
A}$), we would like to prove that
\begin{equation} \label{CS}
 \omega(A^*A)\omega(B^*B)\geq |\omega(B^*A)|^2
\end{equation}
for any two elements $A$ and $B$ in ${\cal A}$. We define a new element
$A':=A\exp(-i{\rm arg}\omega(B^*A))$, so that $|\omega(B^*A)|=\omega(B^*A')$,
and compute
\begin{eqnarray}
 0&\leq& \omega\left( (\sqrt{\omega(B^*B)} A'- \sqrt{\omega(A^{\prime
       *}A')}B)^* (\sqrt{\omega(B^*B)} A'- \sqrt{\omega(A^{\prime
       *}A')}B)\right)\nonumber\\
&=& 2\omega(B^*B)\omega(A^{\prime *}A')- \sqrt{\omega(B^*B)\omega(A^{\prime
    *}A')} \left(\omega(A^{\prime *}B)+\omega(B^*A')\right)\nonumber\\
&=& 2\omega(B^*B)\omega(A^{\prime *}A')- 2\sqrt{\omega(B^*B)\omega(A^{\prime
    *}A')} |\omega(B^*A)|\,.
\end{eqnarray}
Therefore,
\[
|\omega(B^*A)|\leq \sqrt{\omega(B^*B)}\sqrt{\omega(A^{\prime *}A')}=
  \sqrt{\omega(B^*B)\omega(A^*A)}\,.
\]

One can then derive the standard uncertainty relation for basic operators by
applying the Cauchy--Schwarz inequality to
$A=\hat{q}_i-\omega(\hat{q}_i) {\mathbb I}$ and
$B=\hat{p}_j-\omega(\hat{p}_j) {\mathbb I}$: We have $\omega(A^*A)=(\Delta
q_i)^2$, $\omega(B^*B)=(\Delta p_j)^2$, and $\omega(B^*A)$ can be split into
its real part, which equals the covariance $C_{q_ip_j}:=\frac{1}{2}\omega(
\hat{q}_j\hat{p}_i+\hat{p}_i\hat{q}_j)- \omega(\hat{q}_j)
\omega(\hat{p}_i)$, and its imaginary part proportional to the commutator
$[\hat{q}_i,\hat{p}_j]$. The inequality (\ref{CS}) then implies 
\begin{equation}
 (\Delta q_j)^2(\Delta p_i)^2\geq \frac{\hbar^2}{4}\delta_{ij}+C_{q_jp_i}^2\geq
 \frac{\hbar^2}{4}\delta_{ij}\,.
\end{equation}
In the present case, there is a new uncertainty relation
for different components of $p^i$ thanks to the non-zero commutator
(\ref{pp}):
\begin{equation}
 (\Delta p_i)^2(\Delta p_j)^2\geq
 \frac{\hbar^2e^2}{4}(\epsilon_{ijk}\omega(\hat{B}^k))^2
 \delta_{ij}+C_{p_ip_j}^2\geq 
 \frac{\hbar^2e^2}{4}(\epsilon_{ijk}\omega(\hat{B}^k))^2\delta_{ij}\,.
\end{equation}
These relations depend only on commutators, and therefore are
equivalent to those given in \cite{MSS1,BakasLuest,MSS2,MSS3} based on
$*$-products.

At this stage, we note a difference with the $*$-product treatment of
non-associative algebras. When one constructs an analog of a Hilbert-space
representation of wave functions $\psi$ acted on by ${\cal A}$ using a
non-associative $*$-product, one assigns to any $A\in{\cal A}$ a map
$\psi\mapsto A*\psi$ on a set of wave functions $\psi$ instead of the usual
associative action of operators. (The constructions in \cite{MSS2} are more
general and consider also density states.) In deriving the uncertainty
relation, one applies two such multiplications of the form $A*(B*\psi)$. This
product is sensitive to non-associativity, and indeed the derivation of an
uncertainty relation is non-trivial.  In \cite{MSS2}, the problem has been
solved by introducing modified (and associative) composition maps derived but
different from the original algebra product:\footnote{These composition maps
  are important for the construction of states obeying the positivity
  condition \cite{MSS2}.}  $\circ$ is obtained from $(A\circ B)*C=A*(B*C)$ for
all $A,B,C$, and $\bar{\circ}$ from $C*(A\bar{\circ}B)=(C*A)*B$. For the
$*$-product action on states, a Cauchy--Schwarz inequality holds for $\circ$
but not for the original $*$. However, as derived in detail in \cite{MSS2},
the $\circ$-commutator acts by $(\hat{p}_i\circ
\hat{p}_j-\hat{p}_j\circ\hat{p}_i)*\psi= \psi*\hat{K}$, with $\hat{K}$
corresponding to the right-hand side of the commutator (\ref{pp}), but now
acting from the right. Accordingly, the resulting uncertainty relation is not
of the standard form, unless an additional ``symmetry'' condition is imposed
on wave functions, or $\rho*C=C*\rho$ on density states $\rho$.  The general
derivation of the Cauchy--Schwarz inequality, on the other hand, makes use of
products of at most two operators and is not sensitive to
non-associativity. It implies an uncertainty relation that is equivalent to
the one obtained using $*$-products only if the symmetry condition is
imposed. We view this observation as an additional argument that wave
functions should indeed obey the symmetry condition (as already suggested in
\cite{MSS2}).

\subsection{Failure of the GNS construction}

Given an {\em associative} $*$-algebra ${\cal A}$ and a positive linear
functional $\omega$ on it, one can construct a Hilbert-space representation by
making use of the GNS construction. (See for instance
\cite{LocalQuant,ThirringQuantum}.) It is clear that the construction must fail
in the non-associative case because such an algebra cannot act by standard
operator multiplication on a Hilbert space. Nevertheless, it is interesting to
see where exactly the construction breaks down.

In the GNS construction, one starts with the algebra ${\cal A}$ as a linear
space and constructs a Hilbert space from it. Multiplication in the algebra
then implies an action of the algebra on the Hilbert space. In order to derive
the Hilbert space, one introduces a (degenerate) scalar product on ${\cal A}$
by $\langle A|B\rangle:=\omega(A^*B)$ for all $A,B\in{\cal A}$. The scalar
product is positive semidefinite  because $\omega$ is assumed to be a positive
linear functional, but it has a kernel spanned by all $C\in{\cal A}$ for which
$\omega(C^*C)=0$. Assuming the algebra to be associative, the kernel is a
left-ideal in ${\cal A}$ and can be factored out, leaving a linear space with
a positive definite scalar product which can be completed to a Hilbert
space. 

In this last step, associativity is important. In order to show that the
kernel is a left-ideal, one makes use of the Cauchy--Schwarz inequality and
computes (using associativity only at this place in the present paper)
\begin{equation} \label{ideal}
 |\omega((AC)^*(AC))|^2=|\omega(C^*A^*AC)|^2\leq
 \omega(C^*C)\omega((A^*AC)^*A^*AC)=0\,,
\end{equation}
so that $AC$ is in the kernel for any $A\in{\cal A}$ and $C$ in the
kernel. For a non-associative algebra, (\ref{ideal}) is not available and it
is in general impossible to factor out the kernel consistently in order to
obtain a Hilbert space.

It would be possible to obtain a left ideal from the kernel of $\omega$ if all
$C$ in the kernel would be self-adjoint (or anti-selfadjoint).  For an
alternative algebra, we could then proceed as in (\ref{ideal}) thanks to the
Moufang identity
\begin{equation} \label{Moufang1}
 C(AB)C=(CA)(BC)
\end{equation}
which allows us to write
\begin{eqnarray}
 |\omega((AC)^*(AC))|^2&=&|\omega((CA^*)(AC))|^2= |\omega(C(A^*A)C)|^2\\
 & \leq&
 \omega(C^*C)\omega(((A^*A)C)^*((A^*A)C))=0
\end{eqnarray}
if $C^*=\pm C$. A real Hilbert space would follow from the GNS construction if
the algebra could be restricted to only (anti-)self-adjoint
elements. Unfortunately, however, a closed algebra of (anti-)self-adjoint
elements can be obtained only with (anti-)commutative multiplication.

The GNS construction plays an important role in algebraic approaches to
quantum mechanics and quantum field theory because it shows that Hilbert-space
representations do exist. In particular, using all states in a Hilbert-space
representation, one is assured that sufficiently many positive linear
functionals exist on the algebra, allowing one to derive potential measurement
results. A quantum system would not be considered meaningful if it does not
allow sufficiently many states, for instance when
$\omega(\hat{q}_1)=\omega(\hat{q}_2)$ for all states $\omega$. For every point
$(\bar{q}_1,\bar{q}_2,\bar{q}_3;\bar{p}_1,\bar{p}_2,\bar{p}_3)$ in the
classical phase space in which we expect a semiclassical quantum description
to be available, we should require that there is a state $\omega$ such that
$\omega(\hat{q}_i)=\bar{q}_i$ and $\omega(\hat{p}_j)=\bar{p}_j$. The classical
freedom of choosing initial values then remains unrestricted after
quantization.

If sufficiently many states exist, general features of
expectation-value functionals can be employed to derive generic
properties which are independent of which specific representation is
used. For non-associative algebras, we cannot have standard
Hilbert-space representations, and we are not aware of an alternative
version of the GNS construction that could guarantee the existence of
sufficiently many positive linear functionals on the algebra. The
methods of \cite{MSS2} show that states can be constructed with
an action of the algebra given by a $*$-product, and positivity
properties have been demonstrated. However, as shown by the discussion
of uncertainty relations in the preceding section, the general
algebraic results we make use of here agree with those found by
non-associative $*$-products only when the class of states is
restricted by an additional symmetry condition. To the best of our
knowledge, it is not clear whether sufficiently many positive linear
functionals obeying the symmetry condition do exist. In what follows,
we will have to assume that there are such states, some of whose
properties we will be able to derive.

\subsection{Moments}

Without a Hilbert space, we cannot describe states by wave functions. However,
we can use an alternative set of variables which describes a positive linear
functional $\omega$ on ${\cal A}$ in terms of expectation values $\omega(O)$
and moments of the form $\omega((O-\omega(O){\mathbb I})^n)$ for $O$ one of
the basic operators. More generally, covariance parameters in which $\omega$
is applied to products of $O_i-\omega(O_i){\mathbb I}$ for different values of
$i$ are also required. We introduce these variables and determine some of
their algebraic relations after switching to a physics-oriented notation in
which $\omega(A)$ is written as the expectation value $\omega(A)=\langle
A\rangle$ of an operator $A$. These expectation values, by definition, refer
to a state as a positive linear functional on the algebra; they do not require
wave functions or a Hilbert space. Moreover, we will omit explicit insertions
of the unit operator ${\mathbb I}$ and assume that it is understood in
expressions such as $\hat{A}-\langle\hat{A}\rangle$.

In the associative case, the definition of the moment variables is as follows:
\begin{eqnarray} \label{Moments}
\Delta(p_x^{a_1} q_x^{a_2} p_y^{b_1} q_y^{b_2} p_z^{c_1}q_z^{c_2}) &:=& 
\langle ( \hat{p}_x -\langle\hat{p}_x\rangle)^{a_1} (\hat{q}_x
-\langle\hat{q}_x\rangle)^{a_2}\nonumber \\ 
&&\times(\hat{p}_y -\langle\hat{p}_y\rangle)^{b_1}(\hat{q}_y
-\langle\hat{q}_y\rangle)^{b_2}\nonumber \\ 
&&\times(\hat{p}_z -\langle\hat{p}_z\rangle)^{c_1}(\hat{q}_z
-\langle\hat{q}_z\rangle)^{c_2}  \rangle_{\rm Weyl}
\end{eqnarray}
with totally symmetric or Weyl ordering indicated by the subscript
``Weyl.''  Weyl ordering makes sure that we do not count as different
moments which can be obtained from each other by simple applications
of the commutator. Moreover, the moments of Weyl ordered products in
an associative algebra are defined as real numbers. It turns out to be
useful to define them as expectation values of products of the
differences $\hat{A}-\langle\hat{A}\rangle$ as opposed to products
just of basic operators because a semiclassical state can then be
defined as one in which moments of order $a_1+a_2+b_1+b_2+c_1+c_2=:n$
are of the order $O(\hbar^{n/2})$. In this way, one generalizes the
family of Gaussian states in which this order relationship can be
confirmed by an explicit calculation. We make use of the
$\hbar$-orders in our semiclassical equations derived in
Sections~\ref{s:Alg} and \ref{s:Semi}.

For a non-associative algebra, we have to be careful with the order in
which the products are performed. We define the moments by declaring
that products of operators in them are to be evaluated from the left,
that is
\begin{equation}
\Delta(p_xp_yp_z)=\left\langle \left\lbrace (\hat{p}_x
    -\langle\hat{p}_x\rangle)(\hat{p}_y -\langle\hat{p}_y\rangle)\right\rbrace
  (\hat{p}_z -\langle\hat{p}_z\rangle) \right\rangle_{\rm Weyl}\,.
\end{equation}

A bracket on the space of expectation values and moments is defined via the
commutator
\begin{equation} \label{Poisson}
\lbrace  \langle \hat{O}_1 \rangle,\langle \hat{O}_2 \rangle\rbrace
=\frac{\langle [ \hat{O}_1,\hat{O}_2 ]  \rangle }{i\hbar} 
\end{equation}
combined with the Leibniz rule for products of expectation values. For an
associative algebra, this definition gives rise to a Poisson
bracket;\footnote{The resulting Poisson manifold is much larger than the
  classical phase space, and in fact infinite-dimensional owing to infinitely
  many independent moments.} for a non-associative one, the associator is
turned into a non-zero Jacobiator of (\ref{Poisson}).  Evaluating the bracket
on basic variables gives
\begin{eqnarray}
\left\lbrace \langle\hat{q}_i\rangle, \langle\hat{q}_j\rangle\right\rbrace &=& 
\frac{1}{i\hbar}\langle \left [\hat{q}_i,\hat{q}_j \right ]\rangle =0 \\
\left\lbrace \langle\hat{q}_i\rangle, \langle\hat{p}_j\rangle\right\rbrace &=& 
\frac{1}{i\hbar}\langle \left [\hat{q}_i,\hat{p}_j \right ]\rangle =\delta_{ij}
\\ 
\left\lbrace \langle\hat{p}_i\rangle, \langle\hat{p}_j\rangle\right\rbrace &=& 
\frac{1}{i\hbar}\langle \left [\hat{p}_i,\hat{p}_j \right ]\rangle
=e\epsilon_{ijk}\langle \hat{B}^k \rangle\,. 
\end{eqnarray}
For a magnetic field $B^k$ linear in the $q_i$, the right-hand side of the
last relation is a function of basic expectation values, which from now on we
will abbreviate as $q_i=\langle\hat{q}_i\rangle$. For a quadratic function,
such as $B^k(q_i)=C (q_i)^2$ with a constant $C$, we have
$\langle\hat{B}^k\rangle= C\langle\hat{q}_i^2\rangle= C (q_i)^2+
C\Delta(q_i^2)$ with a moment contribution.  In general, if the magnetic
field is non-linear, we may further expand
\begin{equation} \label{Expand}
 \langle \hat{B}^k \rangle=\langle B^k(q_i+(\hat{q}_i-q_i))\rangle= B^k(q_i)+
 \sum_{a,b,c} \frac{1}{a!b!c!} \frac{\partial^{a+b+c}B^k}{\partial
   q_x^a \partial q_y^b\partial q_z^c} \Delta(q_x^aq_y^bq_z^c)
\end{equation}
with a series of moment contributions. There will be an infinite number of
terms if $B$ is non-polynomial. Such an expansion is usually asymptotic and
gives rise to semiclassical or effective equations following the methods of
\cite{EffAc,Karpacz}.

\subsection{Volume uncertainty and uncertainty volume}

Moments are subject to uncertainty relations and cannot be assigned arbitrary
values. For covariances and fluctuations (\ref{Moments}) with
$a_1+a_2+b_1+b_2+c_1+c_2=2$, the standard uncertainty relation follows from
the Cauchy--Schwarz inequality with $\hat{A}=\widehat{\Delta
  O}_1=\hat{O}_1-\langle\hat{O}_1\rangle$ and $\hat{B}=\widehat{\Delta
  O}_2=\hat{O}_2-\langle\hat{O}_2\rangle$ linear in basic operators
$\hat{O}_1$ and $\hat{O}_2$. Moments of higher order are restricted by
uncertainty relations that follow from the Cauchy--Schwarz inequality with
$\hat{A}$ and $\hat{B}$ polynomial in $\widehat{\Delta O}_i$. (See for instace
\cite{Casimir,ClassMoments}.)

A non-zero commutator between two observables $\hat{O}_1$ and
$\hat{O}_2$ provides a lower bound for the product of their
fluctuations $\Delta O_1$ and $\Delta O_2$. For a set of $n$ canonical
pairs $(\hat{q}_i,\hat{p}_i)$, the lower bound of
$\prod_{i=1}^n(\Delta q_i \Delta p_i)\geq (\hbar/2)^n$ may then be
interpreted as an elementary chunk of phase-space volume. For
non-canonical commutators, different lower bounds may be realized for
a subset of the phase-space variables or even the configuration (or
momentum) variables among themselves. For instance, (\ref{pp})
suggests that areas in momentum space have lower bounds given by, for
instance, $\Delta p_x\Delta p_y\geq \frac{1}{2}\hbar e
\langle\hat{B}^z\rangle$, depending on the magnetic field and
therefore, possibly, on the position. A new suggestion, going back to
\cite{DualDoubled} and further analyzed in \cite{MSS2}, is that a
non-zero associator may provide an independent lower bound for triple
products of fluctuations, such as $\Delta p_x\Delta p_y\Delta p_z$ for
(\ref{Asso}). Non-associativity in position space may then,
intriguingly, imply spatial discreteness. (However, even if there is a
lower bound for quantum fluctuations, the relation to discrete
structures is not obvious: In \cite{MomentGUP}, uncertainty relations
have been computed for a discrete system, given by the cotangent space
of a circle, and no lower bound for fluctuations of the discrete
momentum was obtained. As discussed there, such lower bounds could
rather be taken as an indication for extended fundamental objects, as
would be appropriate for lower bounds found in models of string
theory.)

It is not obvious how such uncertainty relations may be derived in a
general way. The Cauchy--Schwarz inequality quite naturally leads to
commutators by expressing the expectation value $\langle A^*B\rangle$
in terms of symmetric and antisymmetric combinations of $A$ and
$B$. It is more difficult to see how the associator might appear in
uncertainty relations as an intrinsic quantity (as opposed to a
quantity derived from the commutator which happens to resemble the
associator). For instance, given a non-trivial uncertainty relation
between momentum components, such as
\[
 (\Delta p_x)^2(\Delta p_y)^2\geq \frac{1}{4}\hbar^2 e^2
 \langle\hat{B}^z\rangle^2\,,
\]
and the standard uncertainty relation between $\Delta q_z$ and $\Delta p_z$, a
magnetic field with $\partial B^z/\partial q_z\not=0$ would imply a
non-trivial lower bound for the triple product
\begin{eqnarray}
 (\Delta p_x)^2(\Delta p_y)^2(\Delta p_z)^2&\geq& \frac{1}{4} \hbar^2e^2
 \langle\hat{B}^z\rangle^2(\Delta p_z)^2\\
&=& \frac{1}{4} \hbar^2e^2 \left(B^z(\langle\hat{q}_j\rangle)^2+
  B^z(\langle\hat{q}_j\rangle)\frac{\partial^2
    B^z(\langle\hat{q}_j\rangle)}{\partial \langle 
    \hat{q}_z\rangle^2} (\Delta q_z)^2+\cdots\right) (\Delta
p_z)^2\nonumber\\
&\geq& \frac{1}{16} \hbar^4 e^2 B^z\frac{\partial^2 B^z}{\partial
  q_z^2}+ \frac{1}{4}\hbar^2e^2 (B^z)^2 (\Delta
p_z)^2+ \cdots \label{D3}
\end{eqnarray} 
However, such an uncertainty relation is neither simple enough to suggest a
universal and state-independent lower bound, nor does it follow directly from
the associator. Moreover, there would be no lower bound for a linear magnetic
field, or a constant associator. (For a semiclassical state, the second term
in (\ref{D3}) would be dominant, so that the inequality would just amount to
the momentum uncertainty relation for $\Delta p_x$ and $\Delta p_y$,
multiplied with an additional factor of $\Delta p_z$ on both sides.)

A direct definition of volume uncertainty would be the uncertainty $\Delta
V=\sqrt{\langle\hat{V}^2\rangle-\langle\hat{V}\rangle}$ of the volume operator
$\hat{V}:=\left((\hat{p}_x\hat{p}_y)\hat{p}_z\right)_{\rm Weyl}$. (The
definition $\hat{V}:=\left(\hat{p}_x(\hat{p}_y\hat{p}_z)\right)_{\rm Weyl}$
would result in the same operator for an alternative algebra.) However, an
uncertainty relation follows from the Cauchy-Schwarz inequality only when
$\Delta V$ is combined with the fluctuation of another observable not
commuting with $\hat{V}$. No universal lower bound for $\Delta V$ itself would
be implied.

One can introduce different quantities which may capture some of the intuition
that may be associated with the notion of ``volume uncertainty.''  For
instance, the quantity $(\widehat{\Delta p}_x\widehat{\Delta
  p}_y)\widehat{\Delta p}_z$ could be related to the associator. In
what follows, we call this triple product of uncertainties the uncertainty
volume, in order to distinguish it from the uncertainty of the volume
operator. (As noted in \cite{MSS2}, the antisymmetrized uncertainty volume is
related to the associator, but it is not clear to us how this quantity may
appear as an upper or lower bound.)

Although the uncertainty volume does appear in some uncertainty relations, it
turns out that it is subject to an upper rather than lower bound by
higher-order uncertainty relations. Choosing
$\hat{A}=\frac{1}{2}(\widehat{\Delta p}_x\widehat{\Delta p}_y+ \widehat{\Delta
  p}_y\widehat{\Delta p}_x)$ and $\hat{B}=\widehat{\Delta p}_z$, one can
compute
\begin{eqnarray}
  \langle\hat{A}^*\hat{A}\rangle &=& \Delta(p_x^2p_y^2)- \frac{1}{4}
  \langle[\widehat{\Delta p}_x,\widehat{\Delta p}_y]^2\rangle+ \frac{1}{6}
  \langle \widehat{\Delta p}_y[[\widehat{\Delta p}_x,\widehat{\Delta
    p}_y],\widehat{\Delta p}_x]- \widehat{\Delta p}_x[[\widehat{\Delta
    p}_x,\widehat{\Delta p}_y],\widehat{\Delta p}_y]\rangle \nonumber\\ 
  &=& \Delta(p_x^2p_y^2)+\frac{e^2\hbar^2}{4}
  \langle \hat{B}^z\rangle+ \frac{e^2\hbar^2}{6} \left\langle \widehat{\Delta
      p}_x 
    \widehat{\partial_yB^z}
   - \widehat{\Delta p}_y
    \widehat{\partial_xB^z}\right\rangle\,.
\end{eqnarray}
(For a linear magnetic field, the last term is zero.) We obtain
$\langle\hat{B}^*\hat{B}\rangle=(\Delta p_z)^2$, as usual, and
$\langle\hat{A}^*\hat{B}\rangle$ contains in its real part the fluctuation
volume:
\begin{equation}
 \langle\hat{A}^*\hat{B}\rangle= \langle(\widehat{\Delta p_x}\widehat{\Delta
   p_y})\widehat{\Delta p_z}\rangle+ \frac{1}{2}\hbar^2e
 \left\langle \widehat{\partial_zB^z}
   \right\rangle- \frac{1}{4}i\hbar e \langle
 \hat{B}^z\widehat{\Delta p_z}+ \widehat{\Delta p_z}\hat{B}^z\rangle\,.
\end{equation}
Therefore, the uncertainty relation for the fluctuation volume $f:=
\langle(\widehat{\Delta p_x}\widehat{\Delta p_y})\widehat{\Delta p_z}\rangle$
is of the form
\begin{equation}
 \left(f+ \frac{1}{2}\hbar^2e \left\langle\widehat{\partial_zB^z}
   \right\rangle\right)^2\leq 
   \Delta(p_x^2p_y^2)\Delta(p_z^2)+\cdots
\end{equation}
However, the associator again does not play a direct role in the derivation.

\section{Algebra of second-order moments}
\label{s:Alg}

We now calculate some of the brackets between second-order moments, providing
characteristic examples in which different features of alternative algebras
appear. These brackets are useful for Hamiltonian equations of motion once the
dynamics is specified, which we will explore in the next section.

\subsection{Application of Moufang identities}

We begin with an example in which the identity (\ref{Moufang1}) (for $A$, $B$
and $C$ linear in the $\hat{p}_i$) plays an important role. For the bracket of
two covariances of different momentum components, we have
\begin{eqnarray}
P_1&:=&\left\lbrace \Delta(p_xp_y), \Delta(p_yp_z)\right\rbrace\\
 &=&\frac{1}{4i\hbar}\langle [ (\hat{p}_x-p_x)(\hat{p}_y-p_y)+(\hat{p}_y-p_y)(\hat{p}_x-p_x),\nonumber \\
&&\ \ \ \ \ \ \ (\hat{p}_y-p_y)(\hat{p}_z-p_z)+(\hat{p}_z-p_z)(\hat{p}_y-p_y) ]\rangle \nonumber \\
&=&\frac{1}{4i\hbar} \langle [ i\hbar e \hat{B}^z + 2
(\hat{p}_y-p_y)(\hat{p}_x-p_x), i\hbar e \hat{B}^x + 2
(\hat{p}_z-p_z)(\hat{p}_y-p_y) ]\rangle  \nonumber
\end{eqnarray}
using the non-zero commutator (\ref{pp}). We continue and write out the
commutator explicitly,
\begin{eqnarray}
P_1 &=& \frac{1}{i\hbar} \langle ((\hat{p}_y-p_y)(\hat{p}_x-p_x))((\hat{p}_z-p_z)(\hat{p}_y-p_y)) \nonumber \\
&&-((\hat{p}_z-p_z)(\hat{p}_y-p_y))((\hat{p}_y-p_y)(\hat{p}_x-p_x))\\
&&+ \frac{1}{2}i\hbar e (\hat{B}^z
(\hat{p}_z-p_z)(\hat{p}_y-p_y)-(\hat{p}_z-p_z)(\hat{p}_y-p_y)\hat{B}^z)\nonumber \\
&&+\frac{1}{2}i\hbar e (-\hat{B}^x
(\hat{p}_y-p_y)(\hat{p}_x-p_x)+(\hat{p}_y-p_y)(\hat{p}_x-p_x)\hat{B}^x)\rangle
\,.\nonumber 
\end{eqnarray}
The Moufang identity can be used in the first line, but not in the second line
in its present form. Two additional applications of commutators bring the
momentum factors of the second line into the form of (\ref{Moufang}):
\begin{eqnarray}
P_1&=& \frac{1}{i\hbar} \langle
((\hat{p}_y-p_y)(\hat{p}_x-p_x))((\hat{p}_z-p_z)(\hat{p}_y-p_y)) \nonumber \\ 
&&- ((\hat{p}_y-p_y)(\hat{p}_z-p_z)-i\hbar
\hat{B}^x)((\hat{p}_x-p_x)(\hat{p}_y-p_y)-i\hbar \hat{B}^z) \nonumber \\ 
&&+ \frac{1}{2}i\hbar e (\hat{B}^z
(\hat{p}_z-p_z)(\hat{p}_y-p_y)-(\hat{p}_z-p_z)(\hat{p}_y-p_y)\hat{B}^z)
\nonumber\\ 
&&+\frac{1}{2}i\hbar e (-\hat{B}^x
(\hat{p}_y-p_y)(\hat{p}_x-p_x)+(\hat{p}_y-p_y)(\hat{p}_x-p_x)\hat{B}^x)\rangle\,.
\end{eqnarray}

Now distributing the second term and using (\ref{Moufang}), and collecting the
middle terms of the first and second terms into a commutator, we obtain
\begin{eqnarray}
P_1 &=& e \left \langle -(\hat{p}_y-p_y)\hat{B}^y (\hat{p}_y-p_y)\right
\rangle \nonumber \\ 
&&+\frac{e}{2}\left \langle \hat{B}^x (\hat{p}_x-p_x)(\hat{p}_y-p_y) +
  (\hat{p}_y-p_y)(\hat{p}_x-p_x)\hat{B}^x\right \rangle \nonumber\\ 
&&+\frac{e}{2} \left\langle \hat{B}^z (\hat{p}_z-p_z)(\hat{p}_y-p_y) +
  (\hat{p}_y-p_y)(\hat{p}_z-p_z)\hat{B}^z\right \rangle\,.\nonumber 
\end{eqnarray}
We can now expand $\langle\hat{B}^i\rangle$ as in (\ref{Expand}) in order
to express this expectation value in terms of moments. If we keep up to
second-order moments for semiclassical equations, we obtain
\begin{eqnarray}
\left\lbrace \Delta(p_xp_y), \Delta(p_yp_z)\right\rbrace &=& -
eB^y\Delta(p_y^2)+eB^x\Delta(p_xp_y)+eB^z\Delta(p_yp_z)\,.\nonumber 
\end{eqnarray}

\subsection{Application of the associator}

Another example in which a combination of commutators and the associator can
be used directly is
\begin{eqnarray}
P_2&:=&\left\lbrace \Delta(p_xq_z), \Delta(p_yp_z)\right\rbrace \label{P2} \\
&=& 
\frac{1}{2i\hbar}
\langle\left[ (\hat{p}_x-p_x)(\hat{q}_z-q_z),(\hat{p}_y-p_y)(\hat{p}_z-p_z)+(\hat{p}_z-p_z)(\hat{p}_y-p_y) \right]\rangle \nonumber \\
&=& \frac{1}{2i\hbar}\langle ((\hat{q}_z-q_z)(\hat{p}_x-p_x))((\hat{p}_y-p_y)(\hat{p}_z-p_z))\nonumber\\
&&-((\hat{p}_y-p_y)(\hat{p}_z-p_z))((\hat{p}_x-p_x)(\hat{q}_z-q_z)) \rangle+
(y\leftrightarrow z \mbox{ only in $p$-terms}) \,.\nonumber
\end{eqnarray}

The goal here is to bring the triple product of $\hat{p}_i$ in the
first term to the form of the second term; for this reason we use the
associator first, and concentrate only on the first term (omitting the
$(\hat{q}_z-q_z)$ term for now):
\begin{equation}
(\hat{p}_x-p_x)((\hat{p}_y-p_y)(\hat{p}_z-p_z))=
((\hat{p}_x-p_x)(\hat{p}_y-p_y))(\hat{p}_z-p_z)+\hbar^2 
e  \widehat{\partial_iB^i}\,. 
\end{equation}
After using the commutator in the parantheses of the first term on the
right-hand side we arrive at
\begin{equation}
(\hat{p}_x-p_x)((\hat{p}_y-p_y)(\hat{p}_z-p_z))=
((\hat{p}_y-p_y)(\hat{p}_x-p_x))(\hat{p}_z-p_z)+i\hbar 
e\hat{B}^z(\hat{p}_z-p_z)+\hbar^2 e   \widehat{\partial_iB^i} \,.
\end{equation}
Once again, we use the associator followed by the commutator, writing
\begin{eqnarray} \label{pppB}
(\hat{p}_x-p_x)((\hat{p}_y-p_y)(\hat{p}_z-p_z))&=&
(\hat{p}_y-p_y)((\hat{p}_z-p_z)(\hat{p}_x-p_x))\\
&&-i\hbar
e(\hat{p}_y-p_y)\hat{B}^y+i\hbar e\hat{B}^z(\hat{p}_z-p_z)+2\hbar^2 e
  \widehat{\partial_iB^i} \,. \nonumber
\end{eqnarray}
Applying this procedure one last time on the very first term in (\ref{pppB})
yields
\begin{equation}
((\hat{p}_y-p_y)(\hat{p}_z-p_z))(\hat{p}_x-p_x)-i\hbar
e(\hat{p}_y-p_y)\hat{B}^y+i\hbar e\hat{B}^z(\hat{p}_z-p_z) 
+3\hbar^2 e  \widehat{\partial_iB^i}\,.
\end{equation}

So far we have looked only at the first term in (\ref{P2}). Observe that we
brought it to the same form as the second term in (\ref{P2}) up to the position
of $(\hat{q}_z-q_z)$ to the left and right, respectively, which can be
combined into a commutator with $(\hat{p}_z-p_z)$ to yield an $i\hbar$. Now
doing the same calculation with the third and fourth term in (\ref{P2}), which
we did not spell out explicitly, yields the bracket wherein the contributions from 
the associator terms drop out 
\begin{eqnarray}
\left\lbrace \Delta(p_xq_z), \Delta(p_yp_z)\right\rbrace &=&
\frac{1}{2}\left \langle
  (\hat{p}_y-p_y)(\hat{p}_x-p_x)+(\hat{p}_x-p_x)(\hat{p}_y-p_y)\right \rangle
\nonumber \\ 
&&+\frac{e}{2} \left \langle \hat{B}^z (\hat{q}_z-q_z)
  (\hat{p}_z-p_z)+(\hat{q}_z-q_z)(\hat{p}_z-p_z)\hat{B}^z\right \rangle
\nonumber \\ 
&&-\frac{e}{2} \left \langle \hat{B}^y (\hat{q}_z-q_z)
  (\hat{p}_y-p_y)+(\hat{q}_z-q_z)(\hat{p}_y-p_y)\hat{B}^y\right \rangle
\nonumber \\ 
&&- \frac{i \hbar e}{2} \langle \hat{B}^z \rangle \,.
\end{eqnarray}
Expanding to second order in moments, we obtain
\begin{eqnarray}
\left\lbrace \Delta(p_xq_z), \Delta(p_yp_z)\right\rbrace &=& \Delta(p_xp_y)+
eB^z\Delta(p_zq_z)- eB^y\Delta(p_yq_z)\nonumber \\
&&-\frac{i\hbar e}{2} \left(B^z+\frac{1}{2}
   \frac{\partial^2B^z}{\partial q_i\partial q_j}\Delta(q_iq_j)\right)+\cdots
  \,.
\end{eqnarray}
Here and in what follows, the dots indicate terms having moments higher than
second order, or terms of order larger than $\hbar$ in a semiclassical state.

\subsection{Application of commutator identities}

In our third example of the brackets it is sufficient to use standard
commutator identities:
\begin{eqnarray}
\left\lbrace \Delta(p_xq_y), \Delta(p_yq_x)\right\rbrace &=&
\frac{1}{i\hbar}\left\langle \left[ (\hat{p}_x-p_x)(\hat{q}_y-q_y),
    (\hat{p}_y-p_y)(\hat{q}_x-q_x) \right] \right\rangle \\  
&=& \langle (\hat{p}_x-p_x) (\hat{q}_x-q_x) -
(\hat{p}_y-p_y) 
(\hat{q}_y-q_y)+e\hat{B}^z (\hat{q}_y-q_y)(\hat{q}_x-q_x)
\rangle\nonumber \\ 
&=& \Delta(p_xq_x) -\Delta(p_yq_y)+ e B^z \Delta(q_xq_y)
\nonumber
\end{eqnarray}
expanded up to second order in moments.

In general, however, one should be careful with the usual identity
$[\hat{A},\hat{B}\hat{C}]= [\hat{A},\hat{B}]\hat{C}+
\hat{B}[\hat{A},\hat{C}]$ when the algebra is not associative, as has
already been pointed out in \cite{MSS2} in the context of Heisenberg
equations of motion, no longer given by a derivation
$[\cdot,\hat{H}]$. In fact, the equation is not valid in general: We
have
\[
 [\hat{A},\hat{B}\hat{C}]= \hat{A}(\hat{B}\hat{C})-(\hat{B}\hat{C})\hat{A}
\]
and 
\[
 [\hat{A},\hat{B}]\hat{C}+ \hat{B}[\hat{A},\hat{C}]= (\hat{A}\hat{B})\hat{C}-
 (\hat{B}\hat{A})\hat{C}+ \hat{B}(\hat{A}\hat{C})- \hat{B}(\hat{C}\hat{A})\,.
\]
The two terms in the middle of the last equation cancel out only when the
multiplication of three given operators is associative. In our example, we
have at most two momentum components, so that this requirement is
satisfied. In general, one can write the difference of the usual two
expressions as a combination of associators:
\begin{equation}
[\hat{A},\hat{B}\hat{C}]-[\hat{A},\hat{B}]\hat{C}+
\hat{B}[\hat{A},\hat{C}]
= -[\hat{A},\hat{B},\hat{C}]- [\hat{B},\hat{C},\hat{A}]-
[\hat{B},\hat{A},\hat{C}]\,.
\end{equation}
For an alternative algebra, the last two terms cancel out and the difference
is just the negative associator $[\hat{A},\hat{B},\hat{C}]$.

\subsection{Brackets}

Having shown a few explicit calculations, we give here a list of some
more brackets of generic type, including their expansions up to
second-order moments:
\begin{eqnarray}
\{\Delta(p_yp_x), \Delta(q_yq_x) \}  &=&- \Delta(p_xq_x) - \Delta(p_yq_y)\\
\{\Delta(p_xq_y), \Delta(q_xq_y) \}  &=& - \Delta(q_y^2) \\
\{\Delta(p_xq_y), \Delta(p_yq_x) \}  &=& \Delta(p_xq_x) - \Delta(p_yq_y) + e
\langle\hat{B}^z (\hat{q}_y-q_y)(\hat{q}_x-q_x)\rangle \nonumber\\
&=& \Delta(p_xq_x) - \Delta(p_yq_y) +eB^z \Delta(q_xq_y)+\cdots\\
\{\Delta(p_xq_y) , \Delta(p_zq_z) \}  &=& - e \langle\hat{B}^y
(\hat{q}_y-q_y)(\hat{q}_z-q_z)\rangle\nonumber \\
 &=& -eB^y \Delta(q_yq_z)+\cdots\\ 
\{\Delta(p_xq_x), \Delta(p_yq_y)\}  &=& e \langle\hat{B}^z
(\hat{q}_x-q_x)(\hat{q}_y-q_y)\rangle\nonumber \\
 &=& eB^z\Delta(q_xq_y)+\cdots\\
\{\Delta(p_xq_z), \Delta(p_yp_z) \}  &=& \Delta(p_xp_y) +\frac{e}{2}
\langle\hat{B}^z
(\hat{q}_z-q_z)(\hat{p}_z-p_z)+(\hat{q}_z-q_z)(\hat{p}_z-p_z)\hat{B}^z
\rangle\nonumber\\ 
&&- \frac{e}{2}
\langle(\hat{q}_z-q_z)\hat{B}^y(\hat{p}_y-p_y)+(\hat{q}_z-q_z)(\hat{p}_y-p_y)
\hat{B}^y\rangle -\frac{i\hbar e}{2}\langle\hat{B}^z\rangle \nonumber\\ 
&=& \Delta(p_xp_y)+ e B^z\Delta(p_zq_z) - eB^y \Delta(p_yq_z)\nonumber \\
&-&\frac{i\hbar e}{2} \left(B^z+\frac{1}{2}
   \frac{\partial^2B^z}{\partial q_i\partial q_j}\Delta(q_iq_j)\right)+\cdots \\
\{\Delta(p_xq_x), \Delta(p_yp_z) \}  &=& \frac{e}{2} \langle\hat{B}^z
(\hat{q}_x-q_x)(\hat{p}_z-p_z)+
(\hat{q}_x-q_x)(\hat{p}_z-p_z)\hat{B}^z\rangle\nonumber\\ 
&&- \frac{e}{2} \langle(\hat{q}_x-q_x)\hat{B}^y(\hat{p}_y-p_y)+
(\hat{q}_x-q_x)(\hat{p}_y-p_y)\hat{B}^y\rangle \nonumber\\ 
&=& eB^z\Delta(p_zq_x)- eB^y\Delta(p_yq_x)+\cdots
\end{eqnarray}

For a non-constant magnetic field, some of the brackets of basic expectation
values and moments are non-zero as well:
\begin{eqnarray}
 \{p_x,\Delta(p_y^2)\} &=& e \langle \hat{B}^z(\hat{p}_y-p_y)+
 (\hat{p}_y-p_y)\hat{B}^z \rangle\\
 &=& 2 e \frac{\partial B^z}{\partial q_i} \Delta(p_yq_i)+\cdots \,.
\end{eqnarray}

\section{Semiclassical dynamics of a charged particle in a magnetic monopole
  density}
\label{s:Semi}

For algebraic states, the dynamics is defined in terms of a flow of positive
linear functionals $\omega_t$, $t\in {\mathbb R}$ on ${\cal A}$ with respect
to a Hamiltonian $H\in {\cal A}$:
\begin{equation} \label{domegadt}
 \frac{{\rm d} \omega_t(O)}{{\rm d}t}:= \frac{1}{i\hbar} \omega_t([O,H])=
 \{\omega_t(O),\omega_t(H)\}
\end{equation}
in terms of the bracket (\ref{Poisson}). This definition agrees with
the standard Schr\"odinger or Heisenberg flow in the case of an
associative algebra of operators represented on a Hilbert space, but
it does not require this additional structure. (It is also insensitive
to the commutator $[\cdot,\hat{H}]$ no longer being a derivation,
which had been noted in \cite{MSS2}.) We can therefore apply it to the
example of a non-associative algebra studied here.

\subsection{General magnetic field}

We first choose the ``free-particle'' Hamiltonian
\begin{eqnarray} \label{Hfree}
\hat{H} = \frac{1}{2m} (\hat{p}_x^2 + \hat{p}_y^2 +\hat{p}_z^2)\,,
\end{eqnarray}
so that we will be considering the motion of a charged particle in a
background magnetic field without additional forces. Interactions between the
charged particle and the magnetic field are represented by the
non-associativity of the algebra or the Jacobiator of the bracket of
expectation values and moments, rather than terms in the Hamiltonian. We
obtain the quantum Hamiltonian as
\begin{eqnarray}
  H_Q:=\langle\hat{H}\rangle = p_x^2
  +p_y^2+p_z^2+\Delta(p_x^2)+\Delta(p_y^2)+\Delta(p_z^2)
\end{eqnarray}
which generates Hamiltonian equations of motion as per (\ref{domegadt}), now
writing $\omega(\hat{H})=\langle\hat{H}\rangle$.

As an example we look at the $x$-components of the equations of motion,
\begin{eqnarray}
\dot{q}_x &=& \frac{1}{m} p_x \\ 
\dot{p}_x &=& \frac{1}{2m}\{p_x ,p_y^2+p_z^2 \} +\frac{1}{2m} \{ p_x , \langle\hat{p}_y^2\rangle - p_y^2 + \langle\hat{p}_z^2\rangle - p_z^2  \} \nonumber \\
&=& \frac{e}{2m} (\langle\hat{p}_y\hat{B}_z + \hat{B}_z \hat{p}_y\rangle -
\langle\hat{p}_z\hat{B}_y + \hat{B}_y \hat{p}_z\rangle ) \,. \label{pxdot}
\end{eqnarray}
In this expression we expand the magnetic field as 
\begin{eqnarray}
\hat{B}^z(\hat{q}) = B^z(q) +(\hat{q}_i-q_i)\frac{\partial B^z}{\partial q_i}
+ \frac{1}{2}
(\hat{q}_j-q_j)(\hat{q}_i-q_i) \frac{\partial^2B^z}{\partial q_i\partial
  q_j}+\cdots  
\end{eqnarray}
and insert it in the first term of (\ref{pxdot}):
\begin{equation}
 \langle\hat{p}_y\hat{B}^z + \hat{B}^z \hat{p}_y\rangle = 2B^z p_y +
 2\frac{\partial B^z}{\partial q_i} \Delta(p_yq_i)+ p_y
 \frac{\partial^2B^z}{\partial q_i\partial q_j} \Delta(q_iq_j)+\cdots\,.
\end{equation}
Here we just added the (vanishing) contribution $\langle p_y (\hat{q}_i -
q_i)\rangle$. Similarly expanding the second term in (\ref{pxdot}) and using
the definitions of moments we get
\begin{eqnarray} \label{max}
m\ddot{q}_x &=& e (B^z v_y - B^y v_z)\\
&& + \frac{e}{m} \frac{\partial
  B^z}{\partial q_i}\Delta(p_yq_i)- \frac{e}{m} \frac{\partial B^y}{\partial
  q_i} \Delta(p_zq_i)+ \frac{e}{2m} \left( p_y \frac{\partial^2B^z}{\partial
    q_i\partial q_j}- p_z \frac{\partial^2B^y}{\partial q_i\partial
    q_j}\right) \Delta(q_iq_j)+\cdots\,. \nonumber
\end{eqnarray}

The first term is the classical Lorentz force. The additional terms are
quantum corrections to the equation of motion, which vanish for a constant
magnetic field. Indeed, it is well-known that a charged particle in a constant
magnetic field can be described by a harmonic-oscillator Hamiltonian, and the
harmonic oscillator does not give rise to quantum corrections in Ehrenfest
equations.

The moments that appear in (\ref{max}) are themselves subject to
dynamical equations of motion with respect to the effective
Hamiltonian.  We have
\begin{eqnarray} \label{momEOM1}
\dot{\Delta}(p_yq_i) &=& 2\Delta(p_yp_x)\delta_{ix} +2\Delta(p_y^2)\delta_{iy}+
2\Delta(p_yp_z)\delta_{iz}-2eB^z\Delta(p_xq_i)\nonumber\\
&& +2eB^x\Delta(p_zq_i)+2eq_i\frac{\partial B^z}{\partial q_j}\Delta(p_xq_j)
+eq_ip_x\frac{\partial^2 B^z}{\partial q_j\partial q_k}\Delta(q_jq_k)\nonumber\\
& &-2eq_i\frac{\partial B^x}{\partial q_j}\Delta(p_zq_j)
-eq_ip_z\frac{\partial^2 B^x}{\partial q_j\partial
  q_k}\Delta(q_jq_k)+\cdots\,.  
\end{eqnarray}
The equation for $\Delta(p_zq_i)$ is analogous to the equation above. The
remaining moment in (\ref{max}) has an equation of motion of the form
\begin{eqnarray} \label{momEOM2}
\dot{\Delta}(q_iq_j)&=&2\Delta(p_xq_i)\delta_{jx} +2\Delta(p_xq_j)\delta_{ix}
+2\Delta(p_yq_i)\delta_{jx} +2\Delta(p_yq_j)\delta_{ix} \nonumber\\
&& +2\Delta(p_zq_i)\delta_{jx} +2\Delta(p_zq_j)\delta_{ix}+\cdots\,.
\end{eqnarray}
For a closed set of equations, we need an equation of motion for moments of
the form $\Delta(p_yp_x)$, which appears in (\ref{momEOM1}).  This calculation
turns out to be more challenging, but it can be handled by using the
associator as well as the defining identities (\ref{alternating}) for an
alternative algebra
\begin{eqnarray}\label{momEQM9}
\dot{\Delta}(p_yp_x)&=& 2e\Bigg[-B^z\Delta(p_x^2)+B^z\Delta(p_y^2)+p_x
\frac{\partial B^z}{\partial q_i}\Delta(p_x q_i)  
+\frac{p_x^2}{2}\frac{\partial^2 B^z}{\partial q_i \partial
  q_j}\Delta(q_iq_j)\nonumber\\  
& &\,\,\,\,\,\,\,\,\,\,-p_y \frac{\partial B^z}{\partial q_i}\Delta(p_y q_i)
 -\frac{p_y^2}{2}\frac{\partial^2 B^z}{\partial q_i \partial q_j}\Delta(q_iq_j)
-B^y\Delta(p_yp_z)+B^x\Delta(p_xp_z)\nonumber \\  
& &\,\,\,\,\,\,\,\,\,\, - p_z\frac{\partial B^y}{\partial q_i}\Delta(p_yq_i) +
p_z\frac{\partial B^x}{\partial q_i}\Delta(p_xq_i)\nonumber \\ 
& &\,\,\,\,\,\,\,\,\,\, -i\hbar  \frac{\partial^2 B^j}{\partial q_i\partial
  q_j}\Delta(p_z q_i)-\frac{i\hbar p_z}{2}  \left(\frac{\partial B^j}{\partial
    q_j} 
+\frac{1}{2}\frac{\partial^3 B^j}{\partial q_j q_i q_k }\Delta(q_i q_k)
\right)\Bigg]+\cdots 
\end{eqnarray}
We now have a closed system of equations for the moments up to second order.

\subsection{Canonical variables in the absence of a magnetic charge density}

In order to test the quantum corrections for a non-constant magnetic field, we
use moment expansions in a derivation of semiclassical equations for the
canonical variables $q_i$ and $\pi_j=m\dot{q}_j+eA_j$ (with
$\{q_i,\pi_j\}=i\hbar$).  These variables can be used only in the absence of a
magnetic charge density, in which case we can compare their dynamics with
(\ref{max}).

In canonical variables, the Hamiltonian operator (\ref{Hfree}) is 
\begin{equation}
 \hat{H}= \frac{1}{2m} (\hat{\pi}-eA)^2= \frac{1}{2m}\delta^{ij}
 (\pi_i-eA_i)(\pi_j-eA_j)\,.
\end{equation}
To second order in moments, it implies a quantum Hamiltonian
\begin{eqnarray}
 H_Q=\langle\hat{H}\rangle &=& \frac{1}{2m} \delta^{ij}\pi_i\pi_j- \frac{e}{m}
 \delta^{ij}\pi_iA_j+ \frac{e^2}{2m} \delta^{ij} A_iA_j\\
&&+ \frac{1}{2m} \delta^{ij} \Delta(\pi_i\pi_j)- \frac{e}{m} \delta^{ik}
 \frac{\partial A_i}{\partial q_j} \Delta(\pi_kq_j) \nonumber\\
&&- \frac{e}{2m}  \delta^{il}
\left((\pi_i-eA_i) \frac{\partial^2A_l}{\partial q_j\partial
    q_k}- e\frac{\partial A_i}{\partial q_j} \frac{\partial A_l}{\partial q_k}
\right) \Delta(q_jq_k) \nonumber
\end{eqnarray}
where $A_i$ is understood as the classical function
$A_i(\langle\hat{q}_j\rangle)$ evaluated at expectation values.

We compute Hamiltonian equations of motion
\begin{eqnarray}
 \dot{q}_i &=& \frac{1}{m} \pi_i- \frac{e}{m} A_i-
 \frac{e}{2m} \frac{\partial^2 A_i}{\partial q_k\partial q_l}
 \Delta(q_kq_l)\\
&=& \frac{1}{m}(\pi_i-e\langle \hat{A}_i\rangle) \nonumber
\end{eqnarray}
and
\begin{eqnarray}
 \dot{\pi}_i &=& \frac{e}{m} \delta^{jk} \pi_j\frac{\partial A_k}{\partial
   q_i}- \frac{e^2}{m} \delta^{jk} A_j\frac{\partial A_k}{\partial q_i}\\
&&+
 \frac{e}{m} \delta^{jk} \frac{\partial^2A_j}{\partial q_i\partial q_l}
 \delta(\pi_kq_l) \nonumber\\
&&+\frac{q}{2m} \delta^{jk}\left( (\pi_j-eA_j) \frac{\partial^3 A_k}{\partial
    q_i\partial q_m\partial q_n}\right. \nonumber\\
&& \left.- e \left(\frac{\partial A_j}{\partial q_i}
    \frac{\partial^2A_k}{\partial q_m\partial q_n}+ \frac{\partial
\dot{\Delta}(p_yp_x)      A_j}{\partial q_m} \frac{\partial^2A_k}{\partial q_i\partial q_n}+
    \frac{\partial A_j}{\partial q_n} \frac{\partial^2A_k}{\partial
      q_i\partial q_m} \right)\right) \Delta(q_mq_n)+\cdots\,. \nonumber
\end{eqnarray}
We will also need the equations of motion for some moments:
\begin{equation}
 \dot{\Delta}(q_mq_n) = \frac{1}{m} \left(\Delta(\pi_mq_n)+
   \Delta(\pi_nq_m)\right) -\frac{e}{m}
 \left(\frac{\partial A_m}{\partial q_l} \Delta(q_lq_n)+
   \frac{\partial A_n}{\partial q_l} \Delta(q_lq_m)\right)+\cdots\,.
\end{equation}

With these results, we can rewrite the Hamiltonian equations of motion as
second-order differential equations for the components $q_i$:
\begin{eqnarray}
 m\ddot{q}_i &=& \frac{e}{m} \pi_j
 \left( \delta^{jk}\delta_{il}\frac{\partial A_k}{\partial q_l}-\frac{\partial A_l}{\partial
     q_j}\right)- \frac{e^2}{m} A_j
 \left( \delta^{jk}\delta_{il}\frac{\partial A_k}{\partial q_l}-\frac{\partial A_l}{\partial
     q_j}\right)\\
&&+ \frac{e}{m} \frac{\partial}{\partial q_m} \left(
  \delta^{jk}\delta_{il}\frac{\partial A_k}{\partial q_l}-\frac{\partial A_l}{\partial 
     q_j}\right) \Delta(\pi_jq_m)\nonumber\\
&&+ \frac{e}{2m} \left((\pi_j-eA_j)
  \frac{\partial^2}{\partial 
  q_m\partial q_n}  
\left(  \delta^{jk}\delta_{il}\frac{\partial A_k}{\partial q_l}-\frac{\partial A_l}{\partial 
     q_j}\right)- e\frac{\partial^2A_j}{\partial q_m\partial q_n} \left(
   \delta^{jk}\delta_{il}\frac{\partial A_k}{\partial q_l}-\frac{\partial A_l}{\partial  
     q_j}\right)\right.\nonumber\\
&&\left. - 2e \frac{\partial A_j}{\partial q_m}
 \frac{\partial}{\partial q_n} \left(  \delta^{jk}\delta_{il}\frac{\partial A_k}{\partial
     q_l}-\frac{\partial A_l}{\partial  
     q_j}\right)\right)  \Delta(q_mq_n)+\cdots\,. \nonumber
\end{eqnarray}
After several simplifications, we can bring this equation into the form
\begin{eqnarray}
 m\ddot{q}_i &=& \frac{e}{m}
 (\pi_j-e\langle\hat{A}_j\rangle) 
 \left( \delta^{jk}\delta_{il}\frac{\partial A_k}{\partial q_l}-\frac{\partial A_l}{\partial
     q_j}\right)\\
&&+ \frac{e}{m} \frac{\partial}{\partial q_m} \left(
  \delta^{jk}\delta_{il}\frac{\partial A_k}{\partial q_l}-\frac{\partial A_l}{\partial 
     q_j}\right) \Delta((\pi_j-eA_j)q_m)\nonumber\\
&&+ \frac{e}{2m} (\pi_j-eA_j) \frac{\partial^2}{\partial
  q_m\partial q_n}  
\left(  \delta^{jk}\delta_{il}\frac{\partial A_k}{\partial q_l}-\frac{\partial A_l}{\partial 
     q_j}\right) \Delta(q_mq_n)+\cdots\nonumber\\
&=& \frac{e}{m}
 (\pi_j-e\langle\hat{A}_j\rangle) 
 \left\langle \delta^{jk}\delta_{il}\frac{\partial A_k}{\partial q_l}-\frac{\partial A_l}{\partial
     q_j}\right\rangle\\
&&+ \frac{e}{m} \frac{\partial}{\partial q_m} \left(\delta^{jk}\delta_{il}
  \frac{\partial A_k}{\partial q_l}-\frac{\partial A_l}{\partial 
     q_j}\right) \Delta((\pi_j-eA_j)q_m)+\cdots\,. \nonumber
\end{eqnarray}
This equation agrees with (\ref{max}), but is valid only in the
absence of a magnetic charge density.

\subsection{Potential and magnetic charge density}

If there is a position-dependent potential in addition to the magnetic field,
the effective Hamiltonian is
\begin{eqnarray}
 H_Q &=& \frac{1}{2m} (p_x^2+p_y^2+p_z^2)+V(q_i)\\
 &&+\frac{1}{2m} (\Delta(p_x^2)+\Delta(p_y^2)+\Delta(p_z^2))+ \frac{1}{2}
 \frac{\partial^2 V}{\partial q_i\partial q_j} \Delta(q_iq_j)+\cdots
\,. \nonumber 
\end{eqnarray}
The potential implies the usual additional terms $-\partial V/\partial q_i$
and $-\frac{1}{2}(\partial^3V/\partial q_i\partial q_j\partial
q_k)\Delta(q_jq_k)$ in the equation of motion for $m\ddot{q}_i$.

\begin{eqnarray} \label{maxpot}
m\ddot{q}_x &=& e (B^z v_y - B^y v_z) -\frac{\partial V}{\partial q_x}\\
&& + \frac{e}{m} \frac{\partial
  B^z}{\partial q_i}\Delta(p_yq_i)- \frac{e}{m} \frac{\partial B^y}{\partial
  q_i} \Delta(p_zq_i)+ \frac{e}{2m} \left( p_y \frac{\partial^2B^z}{\partial
    q_i\partial q_j}- p_z \frac{\partial^2B^y}{\partial q_i\partial
    q_j}\right) \Delta(q_iq_j)\nonumber \\
&& -\frac{1}{2}\frac{\partial^3V}{\partial q_i\partial q_j\partial
q_k}\Delta(q_jq_k)+\cdots \,. \nonumber
\end{eqnarray}

Equations of motion for moments (which appear in the above equation) in this
case are modified as follows:
\begin{eqnarray} \label{momEOM}
\dot{\Delta}(p_yq_i) &=& 2\Delta(p_yp_x)\delta_{ix} +2\Delta(p_y^2)\delta_{iy}+
2\Delta(p_yp_z)\delta_{iz}-2eB^z\Delta(p_xq_i)\nonumber\\ 
&& +2eB^x\Delta(p_zq_i)+2eq_i\frac{\partial B^z}{\partial q_j}\Delta(p_xq_j)
+eq_ip_x\frac{\partial^2 B^z}{\partial q_j\partial
  q_k}\Delta(q_jq_k)\nonumber\\ 
& &-2eq_i\frac{\partial B^x}{\partial q_j}\Delta(p_zq_j)
-eq_ip_z\frac{\partial^2 B^x}{\partial q_j\partial
  q_k}\Delta(q_jq_k)\nonumber\\ 
&& -\frac{1}{2}\frac{\partial^2 V}{\partial q_j\partial
  q_k}\left[\Delta(q_iq_j)\delta_{ky}+\Delta(q_iq_k)\delta_{jy} \right]+\cdots\\ 
\dot{\Delta}(q_iq_j)&=&2\Delta(p_xq_i)\delta_{jx} +2\Delta(p_xq_j)\delta_{ix}
+2\Delta(p_yq_i)\delta_{jx} +2\Delta(p_yq_j)\delta_{ix} \nonumber\\ 
&& +2\Delta(p_zq_i)\delta_{jx} +2\Delta(p_zq_j)\delta_{ix}+\cdots\,.
\end{eqnarray}
For completeness, we also note
\begin{eqnarray}\label{momEQM10}
\dot{\Delta}(p_yp_x)&=& 2e\Bigg[-B^z\Delta(p_x^2)+B^z\Delta(p_y^2)+p_x
\frac{\partial B^z}{\partial q_i}\Delta(p_x q_i)  
+\frac{p_x^2}{2}\frac{\partial^2 B^z}{\partial q_i \partial
  q_j}\Delta(q_iq_j)\nonumber\\  
& &\,\,\,\,\,\,\,\,\,\,-p_y \frac{\partial B^z}{\partial q_i}\Delta(p_y q_i)
-\frac{p_y^2}{2}\frac{\partial^2 B^z}{\partial q_i \partial q_j}\Delta(q_iq_j) 
-B^y\Delta(p_yp_z)+B^x\Delta(p_xp_z)\nonumber \\ 
& &\,\,\,\,\,\,\,\,\,\, - p_z\frac{\partial B^y}{\partial q_i}\Delta(p_yq_i) +
p_z\frac{\partial B^x}{\partial q_i}\Delta(p_xq_i)\nonumber \\ 
& &\,\,\,\,\,\,\,\,\,\, -i\hbar  \frac{\partial^2 B^j}{\partial q_i\partial
  q_j}\Delta(p_z q_i)-\frac{i\hbar p_z}{2}  \left(\frac{\partial B^j}{\partial
    q_j} 
+\frac{1}{2}\frac{\partial^3 B^j}{\partial q_j q_i q_k }\Delta(q_i q_k)
\right)\Bigg] \nonumber \\ 
& &\,\,\,\,\,\,\,\,\,\, -\frac{1}{2}\frac{\partial^2 V}{\partial
    q_i \partial q_j}\left(\Delta(p_yq_i)\delta_{jx}+\Delta(p_yq_j)\delta_{ix}+
\Delta(p_xq_i)\delta_{jy}+\Delta(p_xq_j)\delta_{iy}\right)+\cdots \,.
\end{eqnarray}

As is evident from (\ref{momEOM1}), (\ref{momEOM2}), (\ref{momEQM9}), (\ref{momEOM}) and (\ref{momEQM10}) the
equations of motion for $\Delta(p_yq_i)$ (and $\Delta(p_zq_i)$) and $\Delta(p_yp_x)$ get some
additional terms due to the potential , whereas that for $\Delta(q_iq_j)$
remains the same.

\section{Conclusions}

Taking a pragmatic view that leaves aside existence questions, we have shown
that moment methods are efficient for a derivation of some aspects of
non-associative quantum mechanics, regarding in particular uncertainty
relations and semiclassical equations of motion. This general fact is not
surprising because these algebraic methods are representation independent and
do not require a Hilbert space, a property which is useful in some
quantum-gravity models as well in which such methods had been explored and
developed first.

Still, we did encounter several non-trivial steps in this new application,
which are likely to recur when one tries to extend our results to higher
orders. In several explicit examples, we exploited the existence of Moufang
identities in alternative algebras (which, interestingly, also help to
generalize the axioms of quantum mechanics \cite{OctQM}). Based on the
algebraic relations alone, it seems likely that a more complete version of
formal quantum mechanics can be developed, of which we have given here
semiclassical properties.

Nevertheless, there are several interesting mathematical questions left
open. For instance, while we explicitly used only the antisymmetric associator
$[\hat{p}_1,\hat{p}_2,\hat{p}_3]$ of basic momentum components in our
semiclassical derivations, some higher-order terms would require relationships
for associators of products of momenta. Moufang identities would be available
only if all these associators are totally antisymmetric, that is for an
alternative algebra. The $*$-products of \cite{MSS2}, with the same basic
relations as used here, do not provide an alternative algebra. Our
semiclassical results should still hold in these cases because the associator
of basic momenta is totally antisymmetric, but there may be differences at
higher orders. Properties of moments may therefore allow one to distinguish
between different versions of algebras realizing the basic relations
(\ref{qq})--(\ref{Asso}).

\section*{Acknowledgements}

This work was supported in part by NSF grant PHY-1307408. It is a
pleasure to thank Dieter L\"ust, Peter Schupp and Richard Szabo for
valuable comments on a draft of this article, and Stefan Waldmann for
discussions.


\begin{thebibliography}{10}

\bibitem{Isham}
C.~J.\ Isham,
\newblock Topological and Global Aspects of Quantum Theory,
\newblock In B.~S.\ DeWitt and R.\ Stora, editors, {\em Relativity, Groups and
  Topology II}, 1983,
\newblock Lectures given at the 1983 Les Houches Summer School on Relativity,
  Groups and Topology

\bibitem{Woodhouse}
N.~M.~J.\ Woodhouse,
\newblock {\em Geometric quantization},
\newblock Clarendon (Oxford mathematical monographs), 1992

\bibitem{TopoTwist}
J.-S.\ Park,
\newblock Topological Open $P$-Branes, [hep-th/0012141]

\bibitem{WZWTwist}
C.\ Klimcik and T.\ Strobl,
\newblock WZW-Poisson manifolds,
\newblock {\em J.\ Geom.\ Phys.} 43 (2002) 341--344, [math/0104189]

\bibitem{Twisted}
P.\ Severa and A.\ Weinstein,
\newblock Poisson geometry with a 3-form background,
\newblock {\em Prog.\ Theor.\ Phys.\ Suppl.} 144 (2001) 145--154,
  [math/0107133]

\bibitem{NonGeoNonAss}
R.\ Blumenhagen, A.\ Deser, D.\ L\"ust, E.\ Plauschinn, and F.\ Rennecke,
\newblock Non-geometric fluxes, asymmetric strings and nonassociative geometry,
\newblock {\em J.\ Phys.\ A} 44 (2011) 385401, [arXiv:1106.0316]

\bibitem{MSS1}
D.\ Mylonas, P.\ Schupp, and R.~J\ Szabo,
\newblock Membrane Sigma-Models and Quantization of Non-Geometric Flux
  Backgrounds, [arXiv:1207.0926]

\bibitem{BakasLuest}
I.\ Bakas and D.\ L\"ust,
\newblock 3-Cocycles, Non-Associative Star-Products and the Magnetic Paradigm
  of R-Flux String Vacua, [arXiv:1309.3172]

\bibitem{MSS2}
D.\ Mylonas, P.\ Schupp, and R.~J\ Szabo,
\newblock Non-Geometric Fluxes, Quasi-Hopf Twist Deformations and
  Nonassociative Quantum Mechanics, [arXiv:1312.1621]

\bibitem{MSS3}
D.\ Mylonas, P.\ Schupp, and R.~J\ Szabo,
\newblock Nonassociative geometry and twist deformations in non-geometric
  string theory, [arXiv:1402.7306]

\bibitem{DualDoubled}
D.\ L\"ust,
\newblock T-duality and closed string non-commutative (doubled) geometry,
\newblock {\em JHEP} 1012 (2010) 084, [arXiv:1010.1361]

\bibitem{NonAssGrav}
R.\ Blumenhagen and E.\ Plauschinn,
\newblock Nonassociative Gravity in String Theory?,
\newblock {\em J.\ Phys.\ A} 44 (2011) 015401, [arXiv:1010.1263]

\bibitem{TwistedNonAss}
D.\ L\"ust,
\newblock Twisted Poisson structures and non-commutative/non-associative closed
  string geometry, [arXiv:1205.0100]

\bibitem{NonAssDef}
R.\ Blumenhagen, M.\ Fuchs, F.\ Hassler, D.\ L\"ust, and R.\ Sun,
\newblock Non-associative Deformations of Geometry in Double Field Theory,
  [arXiv:1312.0719]

\bibitem{MagneticCharge}
H.~J.\ Lipkin, W.~I.\ Weisberger, and M.\ Peshkin,
\newblock Magnetic charge quantization and angular momentum,
\newblock {\em Ann.\ Phys.} 53 (1969) 203--214

\bibitem{MonopoleIce}
M.~J.~P.\ Gingras,
\newblock Observing Monopoles in a Magnetic Analog of Ice,
\newblock {\em Science} 326 (2009) 375, [arXiv:1005.3557]

\bibitem{CurrentAlgebra}
K.\ Johnson and F.~E.\ Low,
\newblock Current algebras in a simple model,
\newblock {\em Suppl.\ Prog.\ Theor.\ Phys.} 37 \& 38 (1966) 74--93

\bibitem{CurrentCommutator}
F.\ Buccella, G.\ Veneziano, R.\ Gatto, and S.\ Okubo,
\newblock Necessity of additional unitary-antisymmetric $q$-number terms in the
  commutators of spatial current components,
\newblock {\em Phys.\ Rev.} 149 (1966) 1268--1272

\bibitem{NonAssChiral}
S.~G.\ Jo,
\newblock Commutators in an anomalous non-Abelian chiral gauge theory,
\newblock {\em Phys.\ Lett.\ B} 163 (1985) 353--359

\bibitem{Malcev}
M.\ G\"unaydin and B.\ Zumino,
\newblock Magnetic charge and non-associative algebras,
\newblock In {\em Symposium to honor G.~C.\ Wick}, 1984

\bibitem{Malcev2}
M.\ G\"unaydin and D.\ Minic,
\newblock Nonassociativity, Malcev algebras and string theory,
  [arXiv:1304.0410]

\bibitem{Moufang}
R.\ Moufang,
\newblock Alternativek\"orper und der Satz vom vollst\"andigen Vierseit,
\newblock {\em Abh.\ Math.\ Sem.\ Univ.\ Hamburg} 9 (1933) 207--222

\bibitem{OctQM}
M.\ G\"unaydin, C.\ Piron, and H.\ Ruegg,
\newblock Moufang plane and octonionic quantum mechanics,
\newblock {\em Commun.\ Math.\ Phys.} 61 (1978) 69--85

\bibitem{LocalQuant}
R.\ Haag,
\newblock {\em Local Quantum Physics},
\newblock Springer-Verlag, Berlin, Heidelberg, New York, 1992

\bibitem{ThirringQuantum}
W.\ Thirring,
\newblock {\em Quantum Mathematical Physics},
\newblock Springer, New York, 2002

\bibitem{EffAc}
M.\ Bojowald and A.\ Skirzewski,
\newblock Effective Equations of Motion for Quantum Systems,
\newblock {\em Rev.\ Math.\ Phys.} 18 (2006) 713--745, [math-ph/0511043]

\bibitem{Karpacz}
M.\ Bojowald and A.\ Skirzewski,
\newblock Quantum Gravity and Higher Curvature Actions,
\newblock {\em Int.\ J.\ Geom.\ Meth.\ Mod.\ Phys.} 4 (2007) 25--52,
  [hep-th/0606232],
\newblock Proceedings of ``Current Mathematical Topics in Gravitation and
  Cosmology'' (42nd Karpacz Winter School of Theoretical Physics), Ed.\
  Borowiec, A.\ and Francaviglia, M.

\bibitem{Casimir}
M.\ Bojowald and A.\ Tsobanjan,
\newblock Effective Casimir conditions and group coherent states,
\newblock {\em Class.\ Quantum Grav.} 31 (2014) 115006, [arXiv:1401.5352]

\bibitem{ClassMoments}
D.\ Brizuela,
\newblock Statistical moments for classical and quantum dynamics: formalism and
  generalized uncertainty relations,
\newblock {\em Phys.\ Rev.\ D} 90 (2014) 085027, [arXiv:1410.5776]

\bibitem{MomentGUP}
M.\ Bojowald and A.\ Kempf,
\newblock Generalized uncertainty principles and localization in discrete
  space,
\newblock {\em Phys.\ Rev.\ D} 86 (2012) 085017, [arXiv:1112.0994]

\end{thebibliography}

\end{document}